\documentclass[12pt]{article}
\usepackage{mathrsfs}
\usepackage{fullpage}
\usepackage{amsfonts}
\usepackage{graphicx}
\usepackage{mathrsfs}
\usepackage{amsmath}
\usepackage{amssymb}
\usepackage{float}
\usepackage[round]{natbib}
\usepackage{rotating,color}
\renewcommand\arraystretch{0.8}
\newtheorem{thm}{Theorem}
\newtheorem{lemma}{Lemma}

\def\var{\mathrm {var}}

\def\diag{\mathrm {diag}}
\newcommand{\bm}{\boldsymbol}

\def\tr{\mathrm {tr}}

\def\U{{\bf U}}
\def\A{{\bf A}}
\def\M{{\bf M}}
\def\B{{\bf B}}
\def\Z{{\bm Z}}

\def\R{{\bf R}}
\def\I{{\bf I}}
\def\bmu{{\bm \mu}}
\def\u{{\bf u}}

\def\X{{\bm X}}
\def\x{{\bm x}}

\def\V{{\bm V}}

\def\D{{\bf D}}

\def\tr{\mathrm {tr}}

\def\bmv{\bm \varepsilon}
\def\bth{{\bm\theta}}
\def\bms{{\bm\Sigma}}

\def\cp{\mathop{\rightarrow}\limits^{p}}
\def\cd{\mathop{\rightarrow}\limits^{d}}
\title{Optimal Sign Test for High Dimensional Location Parameters}
\author{Long Feng\\
{\em\small Northeast Normal University}}
\date{}
\begin{document}
\maketitle
\begin{abstract}
This article concerns tests for location parameters in cases where
the data dimension is larger than the sample size. We propose a
family of tests based on the optimality arguments in \citet{51}
under elliptical symmetric. The asymptotic normality of these tests
are established. By maximizing the asymptotic power function, we
propose an uniformly optimal test for all elliptical symmetric
distributions. The optimality is also confirmed by a Monte Carlo
investigation.

\vspace{0.2cm} \noindent{\bf Keywords}: High-dimensional data;
Spatial sign; Uniformly optimal.
\end{abstract}

\section{Introduction}
Testing the population mean vector is a fundamental problem in
statistics. A classical method to deal with this problem is the
famous Hotelling's $T^2$ test. However, it can not work in high
dimensional settings because the sample covariance matrix is not
invertible. With the rapid development of technology, various types
of high-dimensional data have been generated in many areas, such as
internet portals, microarray analysis. By replacing the Mahalanobis
distance by the Euclidean distance, many modified Hotelling's $T^2$
tests for high dimensional data are proposed in many literatures,
such as \citet{3}, \citet{9},\citet{34}, \citet{14}. However, the
statistical performance of the moment-based tests mentioned above
would be degraded when the non-normality is severe, especially for
heavy-tailed distributions.

Many nonparametric methods have been developed, as a reaction to the
Gaussian approach of Hotelling's test, with the objective of
extending to the multivariate context the classical univariate rank
and signed- rank techniques. There are three main groups. One relies
on componentwise rankings \citep{28}, but is not affine invariant.
The second group is based on spatial signs and ranks with the so
called Oja median \citep{25}. Some efforts have been devoted to
extending this type of method to the high dimensional data.
\citet{39} propose a high dimensional spatial sign test by replacing
the scatter matrix with identity matrix. \citet{45} also propose a
scalar-invariant high dimensional sign test for the two sample
location problem. They demonstrate that the multivariate sign and
rank are still very efficient methods in constructing robust test in
high dimension settings. The last group use the concept of
interdirections \citep{29}. In an important work, \citet{50} propose
a class of tests based on interdirections and pseudo-Mahalanobis
ranks. Depending on the score function considered, they allow for
locally asymptotically maximin test at selected densities. However,
to the best of our knowledge,  there are no optimal tests for high
dimensional location parameters.

In this article, we propose an uniformly optimal test for high
dimensional data. Based on the optimality arguments in \citet{51},
we introduce a high dimensional form of the locally and
asymptotically optimal testing procedure. The asymptotic normality
of this class of tests are established. By maximizing the power
function of these tests, we propose an uniformly optimal test for
high dimensional location problem. In the multivariate case, the
optimal score function deeply depends on the underlying
distributions. However, the optimal weighted function for our high
dimensional test is unique. So our proposed test procedure is
uniformly optimal for the elliptical symmetric distributions. We
also derive the asymptotic relative efficiency of our test with
respect to \citet{9}'s test and \citet{39}'s test. It is not
surprised that they are all no less than one for the elliptical
symmetric distributions. And for the heavy tailed distributions,
such as multivariate $t$-distributions or mixture multivariate
normal distributions, our test would perform eventually better than
these two tests. Simulation studies also demonstrate these results.

\section{Uniformly Optimal Test}
\subsection{High Dimensional Weighted Sign tests} Assume
$\{\X_i\}_{i=1}^n$ are i.i.d. random sample from $p$-variate
elliptically symmetric distribution with density function
$\mbox{det}( \Sigma)^{-1/2}g(||{
 \Sigma}^{-1/2}( x-\theta)||)$
where $\theta$ is the symmetry centers and ${\Sigma}$ is the
positive definite symmetric $p\times p$ scatter matrices.  We
consider the following one sample testing problem
\begin{align}\label{hn}
H_0: \bth= 0 ~~~\text{versus}~~~H_1: \bth\not= 0.
\end{align}
When the dimension $p$ is fixed, according to the local asymptotic normality theory \citep{51}, the
form of locally and asymptotically optimal testing procedures for
(\ref{hn}) under specified $\bms$ and $g$ is
\begin{align*}
Q_n=\frac{p}{nc_{p,g}}\sum_{i=1}^n\sum_{j=1}^n\psi_{g}(||\bms^{-1/2}\X_i||)\psi_{g}(||\bms^{-1/2}\X_j||)
U(\bms^{-1/2}\X_i)^TU(\bms^{-1/2}\X_j),
\end{align*}
where $U(\x)=\x/||\x||I(\x\not=\bm 0)$, $\psi_g=-{g^{'}}/{g}$ and
$c_{p,g}$ is a scaled parameter. \citet{50} proposed a class of tests based on interdirections and pseudo-Mahalanobis
ranks which are of the asymptotic form
\begin{align*}
R_n=\frac{2}{n(n-1)}\underset{i<j}{\sum\sum} K(||\bms^{-1/2}\X_i||)K(||\bms^{-1/2}\X_j||)
U(\bms^{-1/2}\X_i)^TU(\bms^{-1/2}\X_j),
\end{align*}
$K(\cdot)$ is a continuous weighted
function. However, the scatter matrix $\bms$
is not available in high dimensional settings. Motivated by
\citet{3} and \citet{9}, we simply replace $\bms$ by $\I_p$ and exclude the same
term in $R_n$. We propose the following generally weighted sign test
statistic:
\begin{align*}
W_n=\frac{2}{n(n-1)}\underset{i<j}{\sum\sum} K(r_i)K(r_j)
U(\X_i)^TU(\X_j),
\end{align*}
where $r_i=||\X_i||$. Let $K(t)=t$, $W_n$ would be the one-sample high dimensional $t$-test
statistic proposed in \citet{9}. Similarly, we can obtain the high
dimensional sign test \citep{39} with $K(t)=1$.  We will determine
the optimal weighted function $K(t)$ in the next section. First, we
propose an asymptotic analysis for $W_n$.

Recently, there are many high dimensional scalar-invariant tests in
literature  \citep{26,34,45,14}. The idea is replacing $\Sigma$ by
its diagonal matrix. And then all the variables have the same scale.
Here we also standardize each variables first by the estimated
diagonal matrix in \citet{45}, which make $W_n$ invariant under the
scale transformation. Details about the scalar-invariant test are given in the appendix. To expedite our discussion, we assume the
diagonal matrix of $\bms$ are known and equal to one without loss of
generality.

The following conditions are needed.
\begin{itemize}
\item[(C1)] $\tr(\bms^4)=o(\tr^2(\bms^2))$ and $\tr(\bms^2)-p=o(n^{-1}p^2)$.
\item[(C2)] $\nu_4=O(\nu_2^2)$ where $\nu_l=E(K^l(r_i))$.
\end{itemize}
The first condition in (C1) is similar to condition (3.8) in
\citet{9}. Obviously, (C1) will hold if all the eigenvalues of
$\bms$ are bounded. The second condition in Condition (C1) is used
to reduce the difference between the module $||\bmv||$ and
$||\bms^{1/2}\bmv||$. Then, we can get an explicit relationship
between the variance of $W_n$ and $\bms$. Condition (C2) is similar
to Assumption 1 in \citet{42} if we choose $K(t)=t^{-1}$.

\begin{thm}
Under Conditions (C1)-(C2) and $H_0$, as $(p,n)\to \infty$,
\begin{align*}
W_n/\sigma_n\cd N(0,1)
\end{align*}
where $\sigma_n^2=2n^{-2}p^{-2}\nu_2^2\tr(\bms^2)$.
\end{thm}

Similar to \citet{39}, we propose the following ratio-consistent
estimator of $\sigma_n^2$
\begin{align*}
\hat{\sigma}_n^2=2n^{-4}\underset{i\not=j}{\sum\sum}K^2(r_i)K^2(r_j)\{U(\X_i)-\bmu_{i,j}\}^TU(\X_j)\{U(\X_j)-\bmu_{i,j}\}^TU(\X_i),
\end{align*}
where $\bmu_{i,j}=\frac{1}{n-2}\sum_{k\not=i,j}U(\X_k)$. And then we
reject the null hypothesis if $W_n/\hat{\sigma}_n>z_{\alpha}$ where
$z_{\alpha}$ is the upper $\alpha$ quantile of $N(0,1)$.

Next, we consider the asymptotic distribution of $W_n$ under the
alternative hypothesis
\begin{itemize}
\item[(C3)] $\bth^T\bth=O(c_0^{-2}\sigma_n)$,
$\bth^T\bms\bth=o(npc_0^{-2}\sigma_n)$ where
$c_0=E\{K(r_i)r_i^{-1}\}$.
\end{itemize}
Condition (C3) require the difference between $\bmu$ and  $0$ is not
large so that the variance of $W_n$ is still asymptotic
$\sigma_n^2$. It can be viewed as a high-dimensional version of the
local alternative hypotheses.
\begin{thm}
Under Conditions (C1)-(C3), as $(p,n)\to \infty$, we have
\[\frac{W_n-c_0^2\bth^T\bth}{\sigma_n}\cd N(0,1).\]
\end{thm}
\subsection{High Dimensional Optimal Sign test}
According to Theorem 1 and 2, the asymptotic power of our weighted
sign test becomes
\begin{align*}
\beta_{\rm
WS}(||\bth||)&=\Phi\left(-z_{\alpha}+\frac{[E\{K(r_i)r_i^{-1}\}]^2}{E\{K^2(r_i)\}}\frac{pn\bth^T\bth}{\sqrt{2\tr({\bms}^2)}}\right).
\end{align*}
The power function of $W_n$ is an increasing function of
$\frac{[E\{K(r_i)r_i^{-1}\}]^2}{E\{K^2(r_i)\}}$. By the Cauchy
inequality, we have
\begin{align*}
\frac{[E\{K(r_i)r_i^{-1}\}]^2}{E\{K^2(r_i)\}}\le
\frac{E\{K^2(r_i)\}E(r_i^{-2})}{E\{K^2(r_i)\}}=E(r_i^{-2}).
\end{align*}
The maximum of $\beta_{{\rm WS}}(||\bth||)$ is $E(r_i^{-2})$ with
maximizer $K(t)=t^{-1}$. Consequently, we propose the following high
dimensional optimal sign test
\begin{align*}
O_n=\frac{2}{n(n-1)}\underset{i<j}{\sum\sum}
r_i^{-1}r_j^{-1}U(\X_i)^TU(\X_j).
\end{align*}
By Condition (C1) and (C3),
$E(r_i^{-2})=E(||\bmv_i||^{-2})(1+o(1))$,
$\bmv_i=\bms^{-1/2}(\X_i-\bmu)$. So the power function of $T_n$ is
\begin{align*}
\beta_{\rm
OS}(||\bth||)&=\Phi\left(-z_{\alpha}+E(||\bmv||^{-2})\frac{pn\bth^T\bth}{\sqrt{2\tr({\bms}^2)}}\right).
\end{align*}
\citet{9} and \citet{39} show that the asymptotic power of their
proposed tests are
\begin{align*}
\beta_{\rm
CQ}(||\bth||)&=\Phi\left(-z_{\alpha}+\frac{np\bth^T\bth}{E(||
\varepsilon||^2)\sqrt{2\tr({\bms}^2)}}\right),\\
\beta_{\rm
SS}(||\bth||)&=\Phi\left(-z_{\alpha}+(E(||\bmv||^{-1}))^2\frac{np\bth^T\bth}{\sqrt{2\tr({\bms}^2)}}\right).
\end{align*}
Thus, the asymptotic relative efficiency of our proposed test with
respect to these two tests are
\begin{align*}
{\rm ARE}({\rm OS}, {\rm
CQ})=&E(||\bmv||^{-2})E(||\bm\varepsilon||^{2})\ge 1\\
{\rm ARE}({\rm OS}, {\rm
SS})=&\frac{E(||\bmv||^{-2})}{\{E(||\bmv||^{-1})\}^2}=1+\frac{\var(||\bmv||^{-1})}{\{E(||\bmv||^{-1})\}^2}
\ge 1.
\end{align*}
Both of the above two equations only hold when
$||\bmv||/E(||\bmv||)\cp 1$.  If $||\bmv||/E(||\bmv||)\cp 1$, these
three tests are asymptotic equivalent. Otherwise, our proposed test
would perform better than the other two tests.

When $\bmv_i \sim N(0,\I_p)$, $||\bmv_i||/\sqrt{p}\cp 1$. Then,
${\rm ARE}({\rm OS}, {\rm CQ})$ and ${\rm ARE}({\rm OS}, {\rm SS})$
are all equal to one.

When $\bmv_i \sim t_p(0, \I_p, v), $ where $ t_p(0, \I_p, v) $ is
the standard $p$-dimensional multivariate $t$ distribution with $ v
$ degrees of freedom, we have
\begin{align*}
{\rm ARE}({\rm OS}, {\rm CQ})=\frac{v}{v-2},~~~~{\rm ARE}({\rm OS},
{\rm SS})=\frac{v\Gamma^2(v/2)}{2\Gamma^2((v+1)/2)}.
\end{align*}
In this case, $\psi_g(t)=(p+v)t/(v+t^2) \to pt^{-1}$ as $t\to
\infty$. So, our uniformly optimal weighted function $K(t)$ would be
consistent with the ``optimal" weighted function $\psi_g(t)$.

When $\bmv_i$ is from the mixtures of two multivariate normal
distributions $MN(\kappa,\sigma,\I_p)$ with density function $
(1-\kappa) f_p(0, \I_p) + \kappa f_p(0,\sigma^2\I_p), $ where $
f_p(;) $ is the density function of $p$-variate multivariate normal
distribution, we have
\begin{align*}
{\rm ARE}({\rm OS}, {\rm
CQ})=(1-\kappa+\kappa/\sigma^2)(1-\kappa+\kappa\sigma^2),~~~~{\rm
ARE}({\rm OS}, {\rm
SS})=\frac{1-\kappa+\kappa/\sigma^2}{(1-\kappa+\kappa/\sigma)^2}.
\end{align*}
As $\sigma^2 \to \infty$, ${\rm ARE}({\rm OS}, {\rm CQ})$ will be
arbitrary large and ${\rm ARE}({\rm OS}, {\rm SS})$ will converge to
$1/(1-\kappa)$. However, in this case,
$\psi_g(t)=\frac{(1-\kappa)t\exp(-t^2/2)+\sigma^{-3}\kappa
t\exp(-t^2/(2\sigma^2))}{(1-\kappa)\exp(-t^2/2)+\sigma^{-3}\kappa
\exp(-t^2/(2\sigma^2))} \to t$ as $t\to \infty$, which is consistent
with \citet{9}'s test. So, $K(t)=\psi_g(t)$ would not be optimal in
such case. Thus, for high dimensional data, a simply extension of
$Q_n$ with $\psi_g(t)$ may not be always the best test.

Table \ref{t1} reports asymptotic relative efficiency between these
three tests under the multivariate $t$-distributions with different
degrees of freedom and mixture normal distributions. Formulas of
asymptotic relative efficiency with these two distributions are
given in the Supplementary Material.

 \begin{table}[ht]
           \centering
           \caption{Asymptotic relative efficiencies with different distributions. }
           \vspace{0.1cm}
      \renewcommand{\arraystretch}{1.1}
     \tabcolsep 4pt
         \begin{tabular}{cccccccccc}\hline \hline
  & {\tiny $t_p(0,\I_p,3)$} & {\tiny $t_p(0,\I_p,4)$} &{\tiny $t_p(0,\I_p,5)$} &{\tiny
  $t_p(0,\I_p,6)$}
  & {\tiny $N(\bm 0, \I_p)$} &{\tiny $MN(0.2,3,\I_p)$} & {\tiny $MN(0.2,10,\I_p)$} &{\tiny $MN(0.8,10,\I_p)$}\\
  {\rm ARE(SS,CQ)} & 2.54 & 1.76 & 1.51 & 1.38  & 1.00 &2.06 &13.98 &6.28\\
  {\rm ARE(OS,CQ)} & 3.00 & 2.00 & 1.67 & 1.50  & 1.00 &2.25 &16.68 &16.68\\
  {\rm ARE(OS,SS)} & 1.18 & 1.13 & 1.11 & 1.09  & 1.00 &1.09 &1.19  &2.65\\\hline \hline
               \end{tabular}\label{t1}\\
               \vspace{0.2cm}
  { $ t_p(0, \Lambda, v) $, $p$-dimensional multivariate $t$ distribution with $ v $ degrees
of freedom and scatter matrix $\Lambda$;
$MN(\kappa,\sigma,\Lambda)$, mixture multivariate normal
distribution with density function $ (1-\kappa) f_p(0, \Lambda) +
\kappa f_p(0,\sigma^2\Lambda), $ where $ f_p(;) $ is the density
function of $p$-variate multivariate normal distribution.}
           \end{table}

\section{Simulation}
Here we report a simulation study designed to evaluate the
performance of the proposed test. All the simulation results are
based on 2,500 replications. We consider the following five
elliptical distributions: (I) $N(\bth, \bms)$; (II) $t_p(\bth, \bms,
3)$; (III) $t_p(\bth, \bms, 4)$; (IV) $MN(0.2,10,\bms)$; (V)
$MN(0.8,10,\bms)$ and two independent component model
$\X_i=\bms^{1/2}\Z_i+\bmu$, $\Z_i=(Z_{i1},\cdots,Z_{ip})$ where (VI)
$Z_{ij}\sim t_3$; (VII) $Z_{ij}\sim 0.8N(0,1)+0.2N(0,100)$. The
scatter matrix is $\bms=(0.5^{|i-j|})$. The sample size is $n=40$
and the dimension is $p=200,400,800$. Under the alternative
hypothesis, two patterns of allocation are considered: (Dense case):
the first $50\%$ components of $\bth$ are zeros; (Sparse case) the
first $95\%$ components of $\bth$ are zeros. And we fixed
$\bth^T\bth/\sqrt{\tr({ \bms})}=0.1$ for the first four scenarios
(I)-(IV) and (VI), and $\bth^T\bth/\sqrt{\tr({ \bms})}=1$ for
scenario (V) and (VII). We compare our proposed test with
\citet{9}'s test and \citet{39}'s test. Table \ref{t2} reports the
empirical sizes and power of these three tests. All these tests can
control the empirical sizes very well. For multivariate normal
distribution and independent component model, the difference between
these three tests are negligible. It is not strange because
$||\bmv||/\sqrt{p}\cp 1$ in this case. Then, the asymptotic relative
efficiency between these tests are all one. But under the non-normal
cases, both \citet{39}'s test and our proposed test performs better
than \citet{9}'s test in all cases. For heavy-tailed distributions,
those direction-based tests will perform better than those
moment-based tests. Furthermore, our proposed test is more powerful
than \citet{39}'s test in these cases, which is consistent with the
asymptotic analysis. Though \citet{39}'s test is very powerful
method, it loses all the information of the module of the
observations. All these results suggest that our proposed test is
very efficient and robust in a wide range of distributions.

 \begin{table}
                     \centering
                     \caption{Empirical sizes and power ($\%$) comparison at 5\% significance under Scenarios (I)-(V)}
                     \vspace{0.1cm}
                \renewcommand{\arraystretch}{0.8}
               \tabcolsep 9pt{
                   \begin{tabular}{cccccccccccccccc}\hline \hline
                   & \multicolumn{3}{c}{Size} && \multicolumn{3}{c}{Dense} &&\multicolumn{3}{c}{Sparse}\\  \cline{2-4} \cline{6-8}\cline{10-12}
    &CQ &SS & OS &&CQ &SS & OS &&CQ &SS & OS\\
 \multicolumn{12}{c}{$(n,p)=(40,200)$} \\ \hline
 (I)   &  5.8 &  6.3 &  6.2  && 74.9 & 76.6 & 76.0  && 81.0 & 83.5 & 82.8\\
 (II)  &  4.5 &  5.7 &  6.2  && 32.4 & 68.2 & 75.3  && 33.6 & 72.9 & 78.7\\
 (III) &  5.1 &  5.9 &  5.7  && 43.1 & 68.9 & 75.2  && 46.3 & 77.4 & 82.3\\
 (IV)  &  6.1 &  7.1 &  6.2  &&  9.0 & 55.1 & 63.7  && 10.3 & 60.6 & 68.9\\
 (V)   &  6.1 &  7.0 &  5.4  && 12.6 & 58.6 & 94.7  && 13.4 & 64.1 & 96.3\\
 (VI)  &  6.6 &  7.3 &  5.4  && 25.1 & 29.7 & 29.5  && 27.4 & 34.0 &
   34.3\\
 (VII) &  4.8 &  5.1 &  4.8  && 34.8 & 38.6 & 39.4  && 40.9 & 45.3 & 45.1\\
   \multicolumn{12}{c}{$(n,p)=(40,400)$} \\ \hline
 (I)   &  5.2 &  6.0 &  5.9  && 78.6 & 80.1 & 79.9  && 80.3 & 82.6 & 82.3\\
 (II)  &  4.3 &  5.1 &  4.7  && 29.7 & 68.1 & 76.9  && 31.9 & 70.7 & 79.4\\
 (III) &  4.9 &  6.0 &  6.6  && 40.8 & 73.7 & 80.5  && 43.1 & 76.6 & 80.9\\
 (IV)  &  5.4 &  6.5 &  5.3  &&  8.3 & 54.5 & 65.3  &&  8.5 & 59.0 & 68.3\\
 (V)   &  4.7 &  6.9 &  5.1  && 10.6 & 57.9 & 95.2  && 10.6 & 59.9 & 94.6\\
 (VI)  &  3.2 &  4.5 &  4.7  && 23.3 & 27.2 & 27.4  && 24.2 & 27.0 &
   26.4\\
 (VII) &  6.0 &  7.0 &  5.8  && 34.8 & 39.9 & 39.7  && 38.4 & 41.4 & 41.9\\
 \multicolumn{12}{c}{$(n,p)=(40,800)$} \\ \hline
 (I)   &  4.2 &  5.8 &  5.4  && 80.7 & 82.4 & 81.5  && 78.4 & 80.5 & 80.1\\
 (II)  &  5.3 &  5.1 &  5.4  && 31.7 & 69.1 & 77.5  && 31.3 & 72.1 & 79.7\\
 (III) &  5.2 &  5.2 &  5.7  && 43.9 & 74.3 & 80.2  && 44.5 & 74.2 & 81.7\\
 (IV)  &  4.1 &  4.7 &  5.5  &&  6.4 & 54.2 & 65.7  &&  7.3 & 57.8 & 68.1\\
 (V)   &  5.9 &  7.0 &  5.0  && 10.3 & 59.9 & 94.8  &&  9.6 & 60.2 &
 94.7\\
(VI)   &  4.3 &  5.1 &  5.3  && 21.3 & 25.5 & 26.4  && 21.7 & 25.8 &
   26.7\\
(VII)  &  4.7 &  5.7 &  5.4  && 36.8 & 41.0 & 40.1  && 36.3 & 40.6 &
40.7
\\\hline \hline
                    \end{tabular}}\label{t2}\\
                    \vspace{0.2cm}
                  CQ, \citet{9}'s test; SS,
                    \citet{39}'s test; OS, our proposed high dimensional uniformly
                    optimal sign test.
                \end{table}

\section{Discussion}
In this paper, we propose a weighted sign test and determine the
``optimal" weight function by maximizing the power function. Our
asymptotic and numerical results together suggest that the proposed
optimal sign test is quite robust and efficient in testing the
population mean vector. This article concerns the one sample
location problem. Testing the equality of two sample locations are
also a very important problem \citep{r21,r3,r4,r12}. In the two
sample problem, the common mean vector is not specified and need to
be estimated. How to extend our method deserves further study.
Furthermore, the proposed test procedure is essentially developed
under the framework of $L_2$-norm-based tests. In another direction,
\citet{r3} and \citet{r25} used the max-norm or thresholding
approach to construct tests rather than the $L_2$-norm. Generally
speaking, the max-norm test is for more sparse and stronger signals
whereas the $L_2$-norm test is for denser but fainter signals.
\citet{f34} also proposed a power-enhancement test based on a
screening technique. Developing a spatial-sign-based test for sparse
signals is of interest in the future study.

\section*{Appendix A: Scalar-invariant test}
 Here we replace $\bms$ in $R_n$ with its diagonal matrix and define the following test statistic
\begin{align*}
T_n=\frac{2}{n(n-1)}\underset{i<j}{\sum\sum}K(||\hat{\D}_{ij}^{-1/2}
\X_i||)K(||\hat{\D}_{ij}^{-1/2}
\X_j||)U(\hat{\D}_{ij}^{-1/2}
\X_i)^TU(\hat{\D}_{ij}^{-1/2}
\X_j),
\end{align*}
where $\hat{\D}_{ij}$ is the corresponding diagonal matrix estimator using leave-two-out sample $\{\X_k\}_{k\not=i,j}$ in \citet{45}. Now, $T_n$ is invariant under scalar transformations $\X_i\to \B\X_i$, $\B=\diag\{b_1^2,\cdots,b_p^2\}$. Define $\R=\D^{-1/2}\bms\D^{-1/2}$ where $\D$ is the diagonal matrix of $\bms$. Now the conditions (C1)-(C3) become
\begin{itemize}
\item[(C$1^{'}$)] $\tr(\R^4)=o(\tr^2(\R^2))$ and $\tr(\R^2)-p=o(n^{-1}p^2)$.
\item[(C$2^{'}$)]  $\tilde{\nu}_4=O(\tilde{\nu}_2^2)$ where $\tilde{\nu}_l=E(K^l(\tilde{r}_i))$ and $\tilde{r}_i=||\D^{-1/2} \X_i||$.
\item[(C$3^{'}$)] $\bth^T\D^{-1}\bth=O(\tilde{c}_0^{-2}\tilde{\sigma}_n)$,
$\bth^T\D^{-1/2}\R\D^{-1/2}\bth=o(np\tilde{c}_0^{-2}\tilde{\sigma}_n)$ where
$\tilde{c}_0=E\{K(\tilde{r}_i)\tilde{r}_i^{-1}\}$ and $\tilde{\sigma}_n^2=2n^{-2}p^{-2}\tilde{\nu}_2^2\tr(\R^2)$.
\end{itemize}
Furthermore, we need another technical condition for the consistency of $\hat{\D}_{ij}$.
\begin{itemize}
\item[(C$4^{'}$)] $n^{-2}p^2/\tr(\R^2)=O(1)$ and $\log(p)=o(n)$.
\end{itemize}

\begin{thm}
Under Conditions (C$1^{'}$)-(C$4^{'}$), as $n,p\to \infty$, we have
\begin{align*}
\frac{T_n-\tilde{c}_0^2\bth^T\D^{-1}\bth}{\tilde{\sigma}_n}\cd N(0,1).
\end{align*}
\end{thm}
Correspondingly, the ratio-consistent estimator of $\tilde{\sigma}_n^2$ is
\begin{align*}
\breve{\sigma}_n^2=&2n^{-4}\underset{i\not=j}{\sum\sum}K^2(||\hat{\D}_{ij}^{-1/2}\X_i||)K^2(||\hat{\D}_{ij}^{-1/2}\X_j||)
\{U(\hat{\D}_{ij}^{-1/2}\X_i)-\tilde{\bmu}_{i,j}\}^TU(\hat{\D}_{ij}^{-1/2}\X_j)\\
&\times\{U(\hat{\D}_{ij}^{-1/2}\X_j)-\tilde{\bmu}_{i,j}\}^T
U(\hat{\D}_{ij}^{-1/2}\X_i),
\end{align*}
where $\tilde{\bmu}_{i,j}=\frac{1}{n-2}\sum_{k\not=i,j}U(\hat{\D}_{ij}^{-1/2}\X_k)$.

So the asymptotic power function of $T_n$ is
\begin{align*}
\beta_{
T_n}(||\bth||)&=\Phi\left(-z_{\alpha}+\frac{[E\{K(\tilde{r}_i)\tilde{r}_i^{-1}\}]^2}{E\{K^2(\tilde{r}_i)\}}
\frac{pn\bth^T\D^{-1}\bth}{\sqrt{2\tr({\R}^2)}}\right).
\end{align*}
By the Cauchy inequality, the optimal weighted function is also $K(t)=t^{-1}$.

\section*{Appendix B: Technical Details}
Define $\U_i=U(\X_i-\bth)$, $\u_i=U(\bmv_i)$, $r_i^{*}=||\bmv_i||$.
First, we restate Lemma 4 in \citet{42}.
\begin{lemma} \label{le1}
Suppose $\u$ are independent identically distributed uniform on the
unit $p$ sphere. For any $p\times p$ symmetric matrix $\M$, we have
\begin{align*}
E(\u^T\M\u)^2=&\{\tr^2(\M)+2\tr(\M^2)\}/(p^2+2p),\\
E(\u^T\M\u)^4=&\{3\tr^2(\M^2)+6\tr(\M^4)\}/\{p(p+2)(p+4)(p+6)\}.
\end{align*}
\end{lemma}

\subsection*{B1: Proof of Theorem 1}

Obviously, $E(W_n)=0$ and
\begin{align*}
\var(W_n)=&\frac{2}{n(n-1)}E\{K^2(r_i)K^2(r_j)(\U_i^T\U_j)^2\}
\end{align*}
Because
$||\X_i||^2=\bmv_i^T\bms\bmv_i=\bmv_i^T\bmv_i+\bmv_i^T(\bms-\I_p)\bmv_i$
and
$E\{\bmv_i^T(\bms-\I_p)\bmv_i\}=E(||\bmv_i||^2)p^{-1}\{\tr(\bms^2)-p\}$,
So $||\X_i||=||\bmv_i||(1+o_p(1))$. Similarly,
$\U_i=\bms^{1/2}\u_i(1+o_p(1))$. Thus,
\begin{align*}
\var(W_n)=&2n^{-2}E\{K^2(r_i^{*})K^2(r_j^{*})(\u_i^T\bms\u_j)^2\}(1+o(1))\\
=&2n^{-2}p^{-2}\nu_2^2\tr(\bms^2)(1+o(1)).
\end{align*}
Thus, we only need to proof the normality of $W_n$. Define
$W_{nk}=\sum_{i=2}^kZ_{ni}$ where
$Z_{ni}=\sum_{j=1}^{i-1}\frac{1}{n(n-1)}\V_i^T\V_j$,
$\V_i=K(r_i)\U_i$. Let $\A=E(\V_i\V_i^T)$. Let
$\mathcal{F}_{n,i}=\sigma\{\V_1,\cdots,\V_i\}$ be the $\sigma$-field
generated by $\{\V_j, j\le i\}$. Obviously, $E(Z_{ni}\mid
\mathcal{F}_{n,i-1})=0$ and it follows that $\{W_{nk},
\mathcal{F}_{n,k}; 2\le k \le n\}$ is a zero mean martingale. The
central limit theorem (Hall and Hyde, 1980) will hold if we can show
\begin{align}\label{clt1}
\frac{\sum_{j=2}^{n}E(Z_{nj}^2\mid\mathcal{F}_{n,j-1})}{\sigma_n^2}\cp
1.
\end{align}
and for any $\epsilon>0$,
\begin{align}\label{clt2}
\sigma_n^{-2}\sum_{j=2}^{n}E\{Z_{nj}^2I(|Z_{nj}|>
\epsilon\sigma_n|)\mid\mathcal{F}_{n,j-1}\}\cp 0.
\end{align}
It can be shown that
\begin{align*}
\sum_{j=2}^{n}E(Z_{nj}^2|\mathcal{F}_{n,j-1})=&\frac{4}{n^2(n-1)^2}\sum_{j=2}^{n}\sum_{i=1}^{j-1}\V_i^T\A \V_i\\
&+\frac{4}{n^2(n-1)^2}\sum_{j=2}^{n}\underset{i_1<i_2}{\sum^{j-1}\sum^{j-1}}\V_{i_1}^T\A\V_{i_2}\\
\doteq& C_{n1}+C_{n2}
\end{align*}
Obviously, $E(C_{n1})=\frac{2}{n(n-1)}\tr(\A^2)=\sigma_n^2(1+o(1))$
by the calculation of $\var(W_n)$. And
$\var(C_{n1})=O(n^{-5})\var((\V_i^T\A\V_i)^2)$. According to Lemma
1, we have $\var((\V_i^T\A\V_i)^2)=O(\tr^2(\A^2)+\tr(\A^4))$. Thus,
by Condition (C1), we have
$\var(C_{n1})=O(n^{-5})\tr^2(\A^2)=o(\sigma_n^4)$. Thus,
$C_{n1}/\sigma_n^2 \cp 1$. Similarly,
$E(C_{n2}^2)=O(n^{-4})\tr(\A^4)=o(\sigma_n^4)$. Then (\ref{clt1})
holds. Next, to proof (\ref{clt2}), by Chebyshev's inequality, we
only need to show
\begin{align*}
E\left\{\sum_{j=2}^{n}E(Z_{nj}^4|\mathcal{F}_{n,j-1})\right\}=o(\sigma_n^4).
\end{align*}
Note that
\begin{align*}
E\left\{\sum_{j=2}^{n}E(Z_{nj}^4|\mathcal{F}_{n,j-1})\right\}=\sum_{j=2}^{n}E(Z_{nj}^4)
=O(n^{-8})\sum_{j=2}^{n}E\left(\sum_{i=1}^{j-1}\V_j^T\V_i\right)^4.
\end{align*}
which can be decomposed as $3Q+P$ where
\begin{align*}
Q=&O(n^{-8})\sum_{j=2}^n
\underset{s<t}{\sum^{j-1}\sum^{j-1}}E(\V_j^T\V_s\V_s^T\V_j\V_j^T\V_t\V_t^T\V_j)\\
P=&O(n^{-8})\sum_{j=2}^n\sum_{i=1}^{j-1}E\{(\V_j^T\V_i)^4\}
\end{align*}
Obviously, $Q=O(n^{-5})E((\V_j^T\A\V_j)^2)=O(n^{-5})\tr^2(\A^2)$ by
Lemma 1 and Condition (C1). Then $Q=o(\sigma_n^4)$. Similarly, we
can show that $P=O(n^{-6})\tr^2(\A^2)=o(\sigma_n^4)$. Here we
complete the proof. \hfill$\Box$

\subsection*{B2: Proof of Theorem 2}

By the Taylor expansion, we have
\begin{align*}
U(\X_i)=\U_i+r_i^{-1}(\I_p-\U_i\U_i^T)\bth+o_p(n^{-1}).
\end{align*}
Thus, taking the same procedure as Theorem 1, we have
\begin{align*}
W_n=&\frac{2}{n(n-1)}\underset{i<j}{\sum\sum}\V_i^T\V_j+\frac{2}{n(n-1)}\underset{i<j}{\sum\sum}K(r_i)r_i^{-1}\V_j^T\bth\\
&+\frac{2}{n(n-1)}\underset{i<j}{\sum\sum}r_i^{-1}r_j^{-1}K(r_i)K(r_j)\bth^T\bth+o_p(\sigma_n)
\end{align*}
And
\begin{align*}
E\left(\frac{2}{n(n-1)}\underset{i<j}{\sum\sum}K(r_i)r_i^{-1}\V_j^T\bth\right)^2=O(n^{-2}p^{-1}c_0^2\bth^T\bms\bth)=o(\sigma_n^2)
\end{align*}
by Condition (C3). Similarly,
\begin{align*}
\frac{2}{n(n-1)}\underset{i<j}{\sum\sum}r_i^{-1}r_j^{-1}K(r_i)K(r_j)\bth^T\bth=c_0^2\bth^T\bth+o_p(\sigma_n).
\end{align*}
Then,
\begin{align*}
W_n=\frac{2}{n(n-1)}\underset{i<j}{\sum\sum}\V_i^T\V_j+c_0^2\bth^T\bth+o_p(\sigma_n).
\end{align*}
According to Theorem 1, we can easily obtain the result. \hfill$\Box$

\subsection*{B3: Consistency of $\hat{\sigma}_n^2$}
Taking the same procedure as the proof of Theorem 2 in \citet{9}, we have
\begin{align*}
\hat{\sigma}_n^2=&2n^{-4}\underset{i\not=j}{\sum\sum}K^2(r_i)K^2(r_j)(\U_i^T\U_j)^2+o_p(\sigma_n^2)\\
=&2n^{-4}\underset{i\not=j}{\sum\sum}(\V_i^T\V_j)^2+o_p(\sigma_n^2),
\end{align*}
by Condition (C3). According to the proof of Theorem 1, we have $E((\V_i^T\V_j)^2)=\tr(\A^2)=p^{-2}\nu_2^2\tr(\bms^2)(1+o(1))$.
So $E(\hat{\sigma}_n^2)=\sigma_n^2(1+o(1))$. And $\var((\V_i^T\V_j)^2)=o(\tr^2{\A^2})$ by Condition (C1) and (C2). Thus, $\var(\hat{\sigma}_n^2)=o(\sigma_n^4)$. So $\hat{\sigma}_n^2/\sigma_n^2 \cp 1$. \hfill$\Box$

\subsection*{B4: Proof of Theorem 3}
By the Tyler's expansion,
\begin{align*}
U(\hat{\D}_{ij}^{-1/2}\X_i)=&\U_i-(\I_p-\U_i\U_i^T)(\hat{\D}_{ij}^{-1/2}-\D^{-1/2})\U_i\\
&+\tilde{r}_i^{-1}(\I_p-\U_i\U_i^T)\D^{-1/2}\bth+o_p(n^{-1}).
\end{align*}
Taking the same procedure as the proof of Theorem 1 in Feng and Sun (2015), by Conditions (C$1^{'}$), (C$2^{'}$) and (C$4^{'}$), we have
\begin{align*}
T_n=&\frac{2}{n(n-1)}\underset{i<j}{\sum\sum}K(\tilde{r}_i)K(\tilde{r}_j)\u_i^T\bms^{1/2}\D^{-1}\bms^{1/2}\u_j\\
&+\frac{2}{n(n-1)}\underset{i<j}{\sum\sum}K(\tilde{r}_i)\tilde{r}_i^{-1}\U_j^T(\I_p-\U_i\U_i^T)\D^{-1/2}\bth\\
&+\frac{2}{n(n-1)}\underset{i<j}{\sum\sum}K(\tilde{r}_j)\tilde{r}_j^{-1}\U_i^T(\I_p-\U_j\U_j^T)\D^{-1/2}\bth\\
&+\frac{2}{n(n-1)}\underset{i<j}{\sum\sum}K(\tilde{r}_i)K(\tilde{r}_j)\tilde{r}_i^{-1}\tilde{r}_j^{-1}\bth^T
\D^{-1/2}(\I_p-\U_i\U_i^T)(\I_p-\U_j\U_j^T)\D^{-1/2}\bth+o_p(n^{-2})\\
\doteq& T_{n1}+T_{n2}+T_{n3}+T_{n4}.
\end{align*}
By the same arguments as the proof of Theorem 1, we have
\begin{align*}
T_{n1}/\tilde{\sigma}_n\cd N(0,1).
\end{align*}
and
\begin{align*}
E(T_{n2}^2)=E(T_{n3}^2)=O(n^{-1}p^{-1}\tilde{c}_0^2\bth^T\D^{-1/2}\R\D^{-1/2}\bth), ~~T_{n_4}=\tilde{c}_0^2\bth^T\D^{-1}\bth+o_p(\tilde{\sigma}_n).
\end{align*}
Thus, by Condition (C$3^{'}$), $(T_n-\tilde{c}_0^2\bth^T\D^{-1}\bth)/\tilde{\sigma}_n\cd N(0,1).$
\hfill$\Box$

\end{document}